\documentclass[%
 reprint,
  superscriptaddress,
 amsmath,amssymb,amsfonts,
 aps,
 prl,
 floatfix,
 longbibliography
]{revtex4-2}

\usepackage{comment}
\usepackage[utf8]{inputenc}
\usepackage[pdftex]{graphicx} \graphicspath{{}}
\usepackage{float} \usepackage{color}
\usepackage[pdftex,colorlinks=true]{hyperref}
\usepackage{subfigure}
\hypersetup{
  colorlinks=true,
  linkcolor=blue,
  citecolor = blue,
  urlcolor=blue,
}

\usepackage{mathtools}

\DeclarePairedDelimiterX\braket[2]{\langle}{\rangle}{#1 \delimsize\vert #2}
\DeclarePairedDelimiterX\expval[3]{\langle}{\rangle}{#1 \delimsize\vert #2  \delimsize\vert #3}
\DeclarePairedDelimiterX\proj[2]{\delimsize\vert#1\rangle}{\langle#2\delimsize\vert}{ }
\usepackage{siunitx}

\usepackage{tikz}
\usetikzlibrary{arrows, arrows.meta}
\usepackage[T1]{fontenc}

\usepackage{tikz}
\definecolor{lime}{HTML}{A6CE39}
\DeclareRobustCommand{\orcidicon}{%
    \begin{tikzpicture}
    \draw[lime, fill=lime] (0,0) 
    circle [radius=0.13] 
    node[white] {{\fontfamily{qag}\selectfont\tiny ID}};
    \draw[white, fill=white] (-0.0625,0.095) 
    circle [radius=0.007];
    \end{tikzpicture}
    \hspace{-2mm}}

\newcommand{\orcidAD}{\href{https://orcid.org/0009-0000-0645-0772}{\orcidicon}}
\newcommand{\orcidJK}{\href{https://orcid.org/0000-0003-0998-9460}{\orcidicon}}
\newcommand{\orcidML}{\href{https://orcid.org/0000-0002-4027-1919}{\orcidicon}}
\newcommand{\orcidGD}{\href{https://orcid.org/0000-0002-7032-2465}{\orcidicon}}
\newcommand{\orcidHK}{\href{https://orcid.org/0000-0003-4527-7671}{\orcidicon}}
\newcommand{\orcidLS}{\href{https://orcid.org/0000-0001-7652-9574}{\orcidicon}}

\usepackage{blindtext}
\begin{document}

\title{Chiral phases and dynamics of dipoles in triangular optical ladders}

\author{Arjo Dasgupta\orcidAD}
\affiliation{Institut f\"ur Theoretische Physik, Leibniz Universit\"at Hannover, Germany}

\author{Mateusz \L\k{a}cki\orcidML} 
\affiliation{Institute of Theoretical Physics, Jagiellonian University in Krakow, ul. Lojasiewicza 11,
	30-348 Krak\'ow, Poland}

\author{Henning Korbmacher\orcidHK}
\affiliation{Institut f\"ur Theoretische Physik, Leibniz Universit\"at Hannover, Germany}

\author{Gustavo A. Dom\'inguez-Castro\orcidGD}
\affiliation{Centro de Nanociencias y Nanotecnolog\'ia, Universidad Nacional Aut\'onoma de M\'exico, Apartado Postal 14, 22800 Ensenada, Baja California, M\'exico}
 
\author{Jakub Zakrzewski\orcidJK}
\affiliation{Institute of Theoretical Physics, Jagiellonian University in Krakow, ul. Lojasiewicza 11,
	30-348 Krak\'ow, Poland}
\affiliation{Mark Kac Complex Systems Research Center,  Jagiellonian University in Krakow, \L{}ojasiewicza 11, 30-348 Krak\'ow, Poland}

\author{Luis Santos\orcidLS}
\affiliation{Institut f\"ur Theoretische Physik, Leibniz Universit\"at Hannover, Germany}


\date{\today}

\begin{abstract}
Dipoles in triangular optical ladders constitute a flexible platform for the study of the interplay between geometric frustration and long-range anisotropic interactions, and in particular for the observation of the spontaneous onset of chirality. Frustration magnifies the effect of the dipolar interactions in itinerant polarized dipolar bosons. As a result, the dipole-induced transition between a chiral superfluid and a non-chiral two-component superfluid may be observed for current state-of-the-art temperatures even for the weak inter-site interaction characterizing magnetic atoms in standard optical lattices. On the other hand, pinned spin-$1/2$ dipoles, which we discuss in the context of polar molecules in two rotational states, realize frustrated dipolar XXZ spin models. By controlling the external electric field strength and orientation, these systems can explore a rich ground-state landscape including chiral and nematic phases, as well as intriguing chiral dynamics.
\end{abstract}

\maketitle



Frustrated many-body lattice systems
are crucial in quantum magnetism~\cite{Balents2010, Lacroix2011}. Prominent examples include antiferromagnets on triangular~\cite{Capriotti1999, Shirata2012} or kagome lattices~\cite{Yan2011, Depenbrock2012}, and one-dimensional~(1D) $J_1$--$J_2$ models~\cite{MikeskaKolezhuk2004}. Frustration suppresses conventional magnetic order and amplifies the role of quantum fluctuations, thereby opening the way to exotic ground states such as quantum spin liquids~\cite{Lhuillier2011}, valence-bond solids~\cite{ReadSachdev1990} and chiral phases~\cite{Hikihara2008}. The latter exhibit a preferred handedness, breaking spatial inversion symmetry. 
In 1D chains, they are characterized by the vector chirality ${\boldsymbol\kappa}_i = {\mathbf S}_i \times {\mathbf S}_{i+1}$, which measures the rotational sense of neighboring spins. An analogous behavior arises in itinerant bosons on frustrated lattices or under artificial gauge fields, which can host chiral superfluids (CSFs), where currents circulate spontaneously~\cite{OrignacGiamarchi2001, Greschner2013, Atala2014, Zaletel2014, Mancini2015, Kock2015, Mishra2015, Greschner2019, Wang2021, HalatiGiamarchi2023, DiLiberto2023, Barbiero2023, Burba2025, Miotto2025, Hasan2025}.  

Ultracold particles in optical lattices and tweezer arrays are an excellent platform for the study of many-body quantum systems. They allow for the realization of models in which lattice geometry, dimensionality, and connectivity can be precisely engineered. In particular, CSFs, resulting from contact interactions and two nonequivalent minima in the Brillouin zone, have been realized in the second Bloch band of $s$-$p$ and hexagonal lattices~\cite{Kock2015, Wang2021}. Chiral currents have been also observed in the presence of synthetic magnetism~\cite{Atala2014, Mancini2015} and in triangular flux ladders~\cite{Li2023}. Frustrated classical magnetism has been simulated in triangular lattices~\cite{Struck2011}, but the study of frustrated quantum magnetism with contact-interacting atoms is limited by weak super-exchange inter-site couplings.



\begin{figure}[t!]
\centering
\includegraphics[width=0.9\columnwidth]{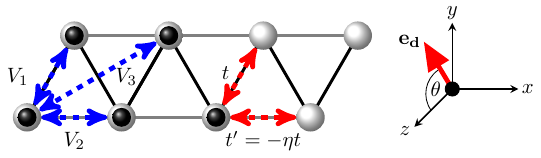}
\caption{Scheme of dipolar bosons on a triangular ladder.}
\label{fig:1}
\end{figure}




\begin{figure*}[t!]
\centering
\includegraphics[width=\textwidth]{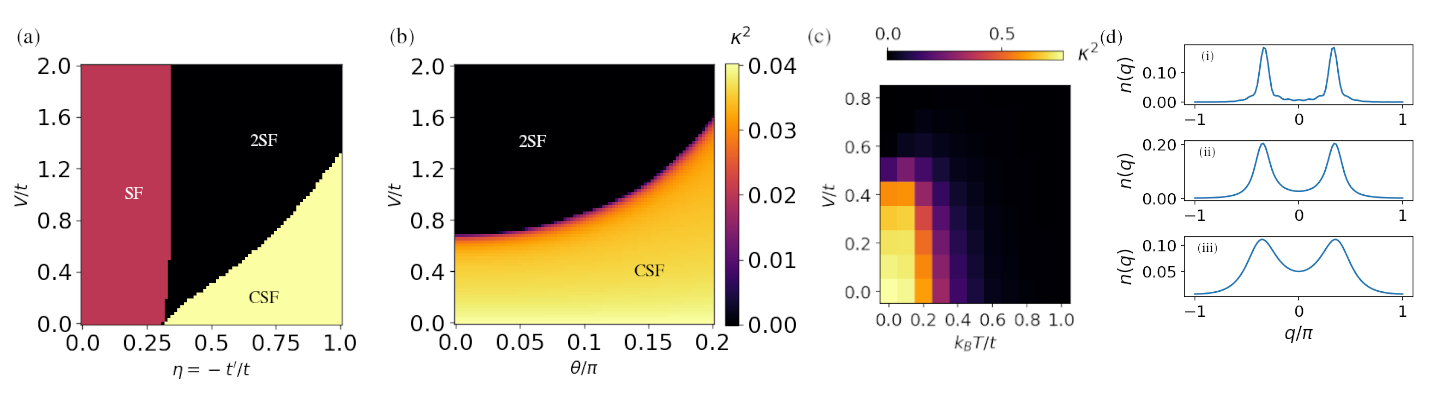}
\caption{(a) Phase diagram for the EBHM~\eqref{eq1} with $U/t=1$, $\theta=0$, and lattice filling $\rho=0.2$ on a ladder with $120$ sites, obtained using DMRG, employing the TeNPy library~\cite{Hauschild2018}, with a maximum bond-dimension of $300$, applying open boundary conditions, and considering up to $3$ particles per site. (b) $\kappa^2$ as a function of $\theta$ and  $V/t$ for the same system at fixed $U/t=1$, and  $\eta=0.75$. (c) $\kappa^2$ as a function of $V/t$ and $k_BT/t$ for $U/t=1$, $\eta=0.6$, and $\rho=1$ for a ladder with $30$ sites. (d) Momentum distribution $n(q)$ in the 2SF at fixed $\eta=0.6$ (i) $k_B  T=0$, $V=0.7t$, (ii) $k_B T=0.6t$,  $V=0.4t$, and (iii) $k_B T=1.8t$, $V=0.2t$.}
\label{fig:2}
\end{figure*}


Lattice experiments using dipolar gases, formed by either magnetic atoms~\cite{Chomaz_2023}, polar molecules~\cite{Moses2017}, or Rydberg atoms~\cite{Browaeys2020}, open interesting possibilities for the study of frustration, since the anisotropic and long-range character of the dipolar interaction results in strong direction-dependent inter-site interactions. Extended Bose-Hubbard models~(EBHMs) formed by polarized itinerant bosons
have been recently engineered in square optical lattices using magnetic atoms~\cite{Baier2016, Su2023}, whereas spin models, formed by pinned dipoles with internal degrees of freedom, have been realized using magnetic atoms~\cite{DePaz2013, Patscheider2020}, polar molecules~\cite{Yan2013, Christakis2023, Ruttley_2025, Hepworth_2025}, 
and Rydberg arrays~\cite{Signoles2021, Scholl2022}.  
Experiments on polar molecules and Rydberg atoms have explored dipolar t-J models~\cite{Carroll2025, Qiao_2025}, including the formation of kinetically-induced bound states in triangular-ladder arrays~\cite{Qiao_2025_b}. Recently, it was proposed that an effective frustrated triangular ladder for magnetic atoms could be realized with two Raman-coupled Zeeman states in a 1D anti-magic wavelength lattice, significantly enhancing the relatively weak nearest-neighbor interactions, which could permit the observation of a CSF-to-supersolid transition~\cite{Miotto2025}.

In this paper, we study the interplay between a frustrated triangular-ladder geometry and the anisotropic long-range dipolar interactions, with a particular emphasis on how that interplay may allow for the observation of chiral phases and dynamics. Geometric frustration magnifies the effect of dipolar interactions in itinerant dipoles allowing for the observation of a dipole-induced transition between a CSF and a non-chiral two-component superfluid for state-of-the-art temperatures~\cite{Mazurenko2017, Su2023, Xu2025} even for the weak inter-site interaction characterizing magnetic atoms in standard lattices~\cite{Su2023}. 
In dipolar spin models, which we discuss in the context of polar molecules, we show that by controlling the strength and orientation of an external electric field, it may be possible to explore a rich ground-state phase diagram including chiral and nematic phases, and  intriguing chiral dynamics.



\paragraph{Itinerant dipoles.--} We first consider itinerant dipolar bosons in a triangular ladder on the $xy$ plane, with legs along $x$~(Fig.~\ref{fig:1}). An external field orients the dipoles along the direction $\boldsymbol{e}_d$ on the $yz$-plane at an angle $\theta$ with the $z$-axis. For deep-enough lattices the system is described by the EBHM:
\begin{eqnarray}\label{eq1}
\hat H &=& -t\sum_i \left(\hat{b}^\dagger_i\hat{b}_{i+1}+\text{H.c.}\right) -t'\sum_i \left(\hat{b}^\dagger_i\hat{b}_{i+2}+\text{H.c.}\right)
\nonumber\\
\!\!\!\!&+&\frac{U}{2}\sum_i \hat{n}_i(\hat{n}_i-1)+ \sum_{i;r>0} V(\mathbf{R}_{i+r}-\mathbf{R}_i) \hat{n}_i \hat{n}_{i+r},
\end{eqnarray}
where $\hat b_i^\dag$ creates bosons at site $i$, and $\hat n_i = \hat b_i^\dag \hat b_i$. Sites with odd~(even) index $i$ belong to the lower~(upper) leg, and are centered at $\mathbf{R}_i=\frac{i}{2}\mathbf{e}_x$~($\mathbf{R}_i=\frac{i}{2}\mathbf{e}_x + \frac{\sqrt{3}}{2}\mathbf{e}_y$), where we set the inter-site spacing to $1$. The hopping rate along the rungs is $t>0$, whereas, crucially, that along the legs is $t'=-\eta t <0$. This may be achieved using Floquet engineering~\cite{Eckardt2017} or Raman-assisted hopping~\cite{Aidelsburger2011}, or alternatively using negative temperature simulations~\cite{Hasan2025}. On-site repulsive interactions, resulting from the interplay of contact and dipolar interactions, are characterized by $U>0$. In contrast, inter-site interactions result purely from the dipole-dipole potential $V(\mathbf{R}) = \frac{V}{|\mathbf{R}|^3} (1-3 (\mathbf{e}_d\cdot \mathbf{R})^2/|\mathbf{R}|^2)$, where $V>0$ determines the dipolar strength. 



\begin{figure*}[t!]
\includegraphics[width=0.85\textwidth]{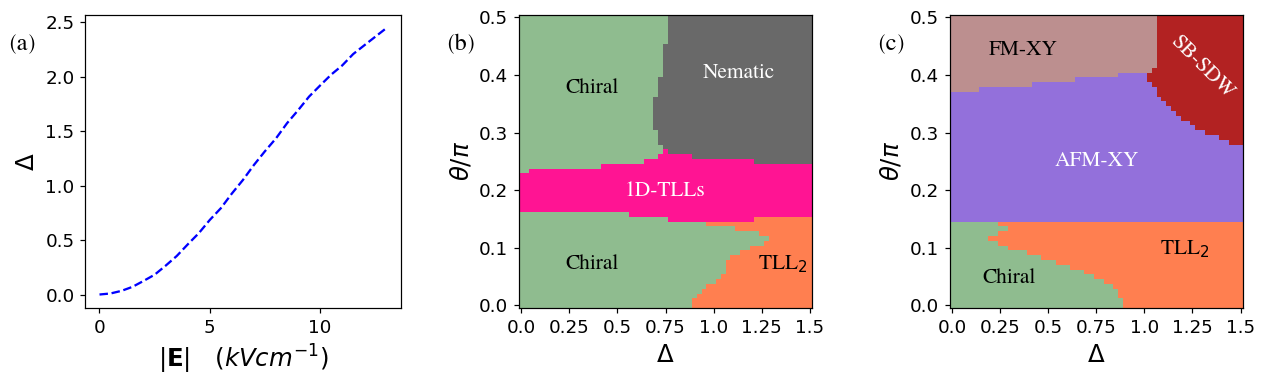}
\caption{(a) Anisotropy $\Delta$  as a function of the electric-field for KRb molecules~\cite{Carroll2025}. (b) Phase diagram of the dipolar XXZ model as a function of $\Delta$ and $\theta$ for dipoles on the $yz$ plane. (c) Same for dipoles on the $xz$ plane. Results obtained for a magnetization $\langle S^z \rangle = -0.4$ using DMRG, employing TeNPy~\cite{Hauschild2018}, with a maximum bond-dimension of $200$, and considering interactions up to $10$ neighbors on a system of $120$ sites with open boundary conditions.}
\label{fig:3}
\end{figure*}




The single-particle dispersion $\varepsilon_q = -2t(\cos(q)-\eta\cos(2q))$ has a single minimum at $q=0$ for $\eta<1/4$, while for $\eta>1/4$ it displays two degenerate minima at $q=\pm Q_0$, with $Q_0= \pm \arccos(1/4\eta)$. For interacting bosons, we may expect that, as long as insulating phases do not occur, the ground-state remains a superfluid~(SF) phase for $\eta<1/4$, with a single peak at $q=0$ in the quasi-momentum distribution 
$n(q) = \sum_{ij} e^{iq(i-j)}\langle \hat b^\dagger_i \hat b_j\rangle$. 
For $\eta>1/4$, interactions may favor one of two phases. In one of them, the bosons occupy just one of the two minima, spontaneously breaking the time-reversal symmetry~($q\rightarrow -q$) of~\eqref{eq1}, forming a superfluid for which 
$n(q)$ has a single peak at $q = \pm Q_0$. This so-called chiral superfluid~(CSF) phase~\cite{Greschner2013, Mishra2015, Greschner2019, HalatiGiamarchi2023, Barbiero2023, Miotto2025} exhibits a vortex-antivortex current distribution around the plaquettes. We characterize this phase 
as that with $\kappa^2 >0$, where $\kappa^2 = \lim_{|i-j|\rightarrow \infty} \langle \hat \kappa_i \hat \kappa_j\rangle$ is the long-range chirality-chirality correlation, with 
$\hat\kappa_j = \frac{1}{2i} (\hat b^\dagger_{j+1} \hat b_j - \hat b^\dagger_{j} \hat b_{j+1})$ the current operator.
Alternatively, the bosons may occupy two equally-populated peaks at $q=\pm Q$, with $Q>0$. This two-component superfluid phase~(2SF)~\cite{Greschner2013, Mishra2015, Greschner2019, HalatiGiamarchi2023} is a Tomonaga-Luttinger liquid~(TLL) with conformal central charge $c=2$ formed by two mutually-incoherent intertwined superfluids. 
Since time-reversal symmetry is not broken, the 2SF presents no chiral currents.

Purely on-site interactions~($V=0$) favor the CSF. In contrast, as illustrated in Fig.~\ref{fig:2}(a) for
$U/t=1$, $\theta=0$, and lattice filling $\rho=0.2$~(results obtained using DMRG, employing the TeNPy library~\cite{Hauschild2018}, for $120$ sites with open boundary conditions),  repulsive dipolar interactions induce a CSF-2SF transition at a critical $V/t$~\cite{SM}.
This may be qualitatively understood 
from a mean-field approach in 
momentum space, $\hat b_q \to \alpha_q$, which results in the mean-field energy: 
\begin{equation}
\label{mean_field}
E[\{\alpha_q\}] = \sum_q \varepsilon_q \alpha_q + \frac{\rho}{2} \sum_q \tilde V_q \, \Big|\sum_k \alpha^*_{k+q}\alpha_k\Big|^2,
\end{equation}
where the momentum distribution $|\alpha_q|^2$ is normalized to $1$. The momentum-dependent interactions $\tilde V_q = U + 2\sum_{r>0} V_{r}\cos(qr) $ favor pairs of peaks at $\pm Q$, which minimize $\tilde V_{2Q}$, and hence the 2SF. 
It is clear that the inherent anisotropy of the dipolar interaction may play a crucial role here. Indeed, the chiral character may be controlled by the dipole orientation. Tilting the dipoles towards the $y$-axis~(increasing $\theta$), reduces the nearest-neighbor~(NN) interaction $V\left(1-\frac{9}{4} \sin^2\theta\right)$, while keeping the next-to-NN interactions fixed at $V$. This may drive a 2SF-CSF transition at fixed $V/t$ and $\eta$, by tuning $\theta$, see  Fig.~\ref{fig:2}(b). 

Interestingly, whereas 
large $V/t$, well over $1$, is typically necessary to reveal dipolar effects, such as density waves~\cite{Su2023} or Haldane insulators~\cite{Dallatorre2006}, the critical $V/t$ for observing the dipole-induced CSF-2SF transition
may be very small, approaching zero at the vicinity of $\eta=1/4$, as a result of the flat band dispersion at $q=0$ induced by frustration~\cite{Mishra2015}.
This is particularly interesting for experiments with magnetic atoms~\cite{Baier2016, Su2023}, for which the inter-site interaction is relatively weak, e.g. $V/h\simeq 30$Hz for erbium atoms at an inter-site separation of $266$nm. Hence, 
dipolar effects at $V/t\sim 5$ could be observed only at the expense of severely reducing the hopping rate $t/h$ down to few Hz, which results in problems related to lattice stability and finite temperature.
Crucially, working at much lower $V/t$, and hence larger $t$, eases the constraints related to finite temperature effects. E.g. working with $V/t=0.5$ in erbium gases in standard optical lattices would demand $t/h\simeq 60$Hz, over one order of magnitude more than in Ref.~\cite{Su2023}. 
The mean-field picture of Eq.~\eqref{mean_field} reveals that the CSF and 2SF phases are separated by an energy gap $\Delta_{CSF-2SF} \sim \rho \tilde V_{2Q}$. For low filling-factors $\rho\ll 1$, this implies $\Delta_{CSF-2SF}\ll t$, limiting the observation of the dipole-induced CSF-2SF transition to temperatures $T\ll t/k_B$, with $k_B$ the Boltzmann constant.
Fortunately, at larger $\rho$ the critical temperature may be significantly higher, as illustrated in Fig.~\ref{fig:2}(c) for  $\rho=1$, $U/t=1$, $\eta=0.6$, and $\theta=0$, for a system of $L=30$ sites~\cite{3body}. We use imaginary time-evolution of purification matrix product states~\cite{Verstraete2004, Hauschild2018a} to obtain a thermal density matrix which describes the system at a finite $T$. The measured $\kappa^2$ indicates that, as discussed above, chirality may be destroyed by both thermal and dipolar effects, but 
realistically reachable temperatures of $k_B T\sim 0.3 t$~\cite{Mazurenko2017, Xu2025} are enough to reveal the dipole-induced destruction of chirality at $V/t\sim 0.5$.  

Finally, note that the resulting thermally-induced non-chiral phase still presents two well-resolved quasi-momentum peaks even at $k_B T>t$ (see Fig.~\ref{fig:2}(d)).



\paragraph{Pinned dipolar spins.--}
Dipolar gases offer an alternative scenario for the onset of chirality without the need of Floquet engineering, which we discuss in the context of polar molecules.
We consider a triangular ladder as that of Fig.~\ref{fig:1}~(formed by an optical lattice or a tweezer array) with each site occupied by a single pinned polar molecule. We assume an external electric field $\mathbf{E}$ along the direction $\mathbf{e}_d$ on the $yz$-plane, forming an angle $\theta$ with the $z$ axis~($\mathbf{e}_d$ sets the quantization axis). As in recent KRb experiments~\cite{Carroll2025} we consider the pseudo-spin-$1/2$ system given by the two lowest dressed rotational states with zero rotational angular momentum projection along the quantization axis~($|0,0\rangle' \equiv |\!\downarrow\rangle$ and $|1,0\rangle' \equiv |\!\uparrow\rangle$). 
Dipole-dipole interactions result in spin-exchange and Ising interactions~\cite{Gorshkov2011}, well described by an extended version (beyond next-to-NNs) of the anisotropic $J_1$-$J_2$ Hamiltonian:
\begin{equation}
\!\!\hat H \!=\! \sum_{i, r>0} \!
J_r \! \left [\frac{1}{2} \left(
\hat S^+_i\hat S^-_{i+r} \!+\!\hat S^-_i\hat S^+_{i+r} \right) \!+\! \Delta \hat S^z_i \hat S^z_{i+r}
\right ],
\end{equation}
where $\hat S^{\pm, z}_i$ are the spin-$1/2$ operators for the molecule at site $i$, $J_{2j+1} = \frac{V}{(j^2-j+1)^{3/2}} \left(1-\frac{9\sin^2\theta}{4(j^2-j+1)}\right)$, and 
$J_{2j} = V/j^3$. The exchange anisotropy $\Delta$ is determined by $|\mathbf{E}|$, see Fig.~\ref{fig:3}~(a) for the case of KRb molecules~\cite{Carroll2025}. 
For all $\theta$ values, the spin model is frustrated and the magnon dispersion at $\Delta=0$, 
$\epsilon_q=\sum_{r>0} J_r \cos(qr)$, 
presents two non-zero minima in the Brillouin zone. 

Fig.~\ref{fig:3}(b) shows the ground-state phase diagram for a magnetization $\langle \hat S^z \rangle = -0.4$~\cite{SM}.
Dominant exchange~(small-enough $\Delta$) results in a chiral phase~\cite{Nersesyan1998}~(a magnon CSF), similar to that reported in 1D 
multiferroic cuprates~\cite{Furukawa2010}. In the spin model, the chiral phase is characterized by a non-zero chirality
$\langle \hat \kappa_i \rangle$, with 
$\hat\kappa_i = \left ( \hat{\mathbf S}_i \times \hat{\mathbf S}_{i+1} \right )^z$. 
As with the EBHM, we identify the chiral phase as that with $\kappa^2>0$, where again 
$\kappa^2 = \lim_{|i-j|\rightarrow \infty} \langle \hat \kappa_i \hat \kappa_j\rangle$. For small enough $\theta<\arcsin(2/3)$, 
all Ising interactions $J_r \Delta$ are repulsive, and hence at a critical $\Delta$~(which plays a similar role as $V/t$ in the EBHM) there is 
a transition from the chiral to the non-chiral two component TLL~(TLL$_2$ phase~\cite{Hikihara2010}), which may be considered a magnon 2SF.  
In the vicinity of $\theta \sim \arcsin(2/3)$, 
$J_1 \simeq 0$, causing the two legs to decouple, leading to a regime of two separate TLLs in each leg~(1D-TLLs), for which the inter-leg correlations 
$\langle S_i^+ S_{i+2n-1}^-\rangle \simeq 0$. For larger $\theta>\arcsin(2/3)$, the chiral phase is recovered at small $\Delta$, but it does not transition into a TLL$_2$ when $\Delta$ grows. In this regime, $J_1<0$ and $J_{r>1}>0$, resulting at a critical $\Delta$ in the formation of NN magnon pairs which delocalize along the legs via resonant next-to-NN exchange. The chiral phase thus transitions into a quadrupolar nematic phase~\cite{Hikihara2008, Sudan2009, Heidrich-Meisner2010}, i.e. a magnon pair-superfluid, characterized by exponentially decaying $C(r) \equiv \langle \hat S_j^+ \hat S_{j+r}^- \rangle$, and polynomially-decaying  $D(r)=\langle \hat S_j^{+}\hat S_{j+1}^{+} \hat S_{j+r}^{-}\hat S_{j+r+1}^{+} \rangle$.


The anisotropy of the dipolar interactions allows for the exploration of markedly different phase diagrams by merely changing the dipole orientation, as illustrated in Fig.~\ref{fig:3}(c)~\cite{SM} for $\langle S^z \rangle = -0.4$ and dipoles on the $xz$ plane forming an angle $\theta$ with the $z$ axis. In that case, $J_{2j-1} = \frac{V}{(j^2-j+1)^{3/2}} \left(1-\frac{3(j-1/2)^2\sin^2(\theta)}{(j^2-j+1)}\right)$ and $J_{2j} = \frac{V}{j^3}\left(1-3\sin^2\theta\right)$.
For $\theta<\theta_0\simeq 0.16\pi$, the 
system is frustrated, experiencing a chiral-to-TLL$_2$ transition at a critical $\Delta$. 
For $\theta> \theta_0$, the system
is unfrustrated~(single minimum in $\epsilon_q$), and at low enough $\Delta$ may be either in an antiferromagnetic-XY~(AFM-XY) or ferromagnetic-XY~(FM-XY) phase, i.e. magnon TLLs with quasimomentum peak at $q=\pi$ or $q=0$. For large-enough $\theta$ and $\Delta$, the attractive $J_{2j}<0$ results in the formation of a self-bound spin-density wave~(SB-SDW), i.e. a self-bound magnon domain at one of the two legs. 
Finally, we note that other orientation planes different than $yz$ or $xz$ break the sublattice symmetry, allowing for the exploration of dimerized dipolar spin-$1/2$ chains~\cite{Ueda2014}.



\begin{figure}[t!]
\includegraphics[width=0.9\columnwidth]{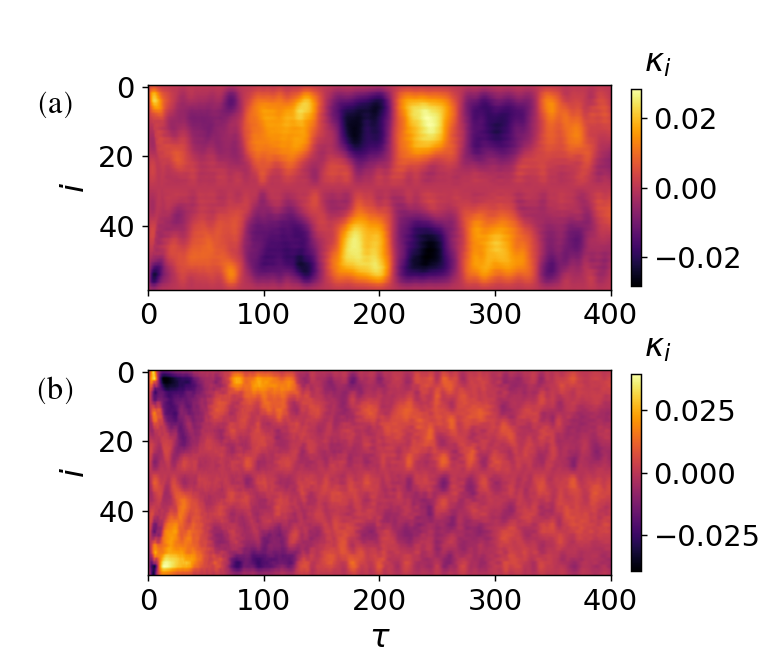}
\caption{Local chirality $\langle \hat \kappa_i\rangle$ as a function of time $\tau$~(in units of $1/V$) for dipoles on the $yz$ plane at an angle $\theta=0.15\pi$ with the $z$-axis, $\langle \hat S^z \rangle = -0.4$, and $\Delta=0.05$~(a) and $1.5$~(b). The initial state is that of two independent magnon TLLs in each leg. The calculations were performed using TDVP, employing TeNPy~\cite{Hauschild2018}, for $60$ sites with open-boundary conditions, keeping a maximum bond-dimension of $300$.}
\label{fig:4}
\end{figure}


\paragraph{Chiral Dynamics.--}
Dipoles in triangular ladders open additional interesting possibilities in what concerns non-equilibrium dynamics, since the system parameters, and even the lattice geometry, may be easily changed in real time. This may be employed to investigate the onset of out-of-equilibrium chiral dynamics from an initial non-chiral state. Let us consider in particular, the case of polar molecules in a tweezer array. The legs of the array are initially separated at enough distance to create two separate magnon TLLs, with magnetization $\langle \hat S^z \rangle = -0.4$. The electric field strength is chosen such that $\Delta=0.05$, and oriented on the $yz$ plane forming an angle $\theta=0.15\pi$ with the $z$ axis~(within the chiral phase in Fig.~\ref{fig:3}(b)). At time $\tau=0$, the two legs are approached forming an equilateral triangular ladder. As shown in Fig.~\ref{fig:4}(a), the initially non-chiral system undergoes a local spontaneous breaking of the chiral symmetry that results in the formation of 
domains which alternate their chirality in time. By tuning the electric field such that $\Delta=1.5$, we can instead choose a system with a ground-state which preserves time-reversal symmetry. The dynamics in this case does not show a clear development of chirality~ (Fig.~\ref{fig:4}(b)).



\paragraph{Conclusions.--}
Dipolar bosons in triangular ladders constitute an interesting platform to study the interplay of 
geometric frustration and long-range anisotropic interactions. 
For mobile particles, frustration resulting from negative in-leg hopping magnifies the effect of dipolar interactions. The dipole-induced CSF-2SF transition may be observed even at very small ratios $V/t < 1$ for experimentally reachable temperatures. This is particularly interesting for experiments with magnetic atoms, where $V$ is relatively small. 
Furthermore, in dipolar spin models, frustration is induced by the dipolar interaction itself. Focusing on the implementation with polar molecules, we have shown that chiral magnon phases, as well as other intriguing phases, including nematic (bound bi-magnon superfluids) and two-component magnon superfluids (TLL$_2$) may be easily realized by properly setting the orientation and magnitude of the applied electric field. 
Finally, let us note that chirality may be experimentally revealed using the dimerization technique recently employed in Ref.~\cite{Impertro2024}. In dipolar gases, due to inter-site interactions, isolating a rung of the triangular ladders demands either selectively removing the particles at the neighboring rungs or physically separating the rungs (in accordion lattices or tweezer arrays). Once the rungs are isolated, currents may be determined from the measurement of the inter-site imbalance of the density~(Hubbard model) or the magnetization~(spin model).  In addition to this method, the three superfluid phases in the model of itinerant bosons can also be distinguished through their momentum distribution, which may be imaged in time-of-flight experiments.


\acknowledgments
We thank L. Barbiero and A. M. Rey for enlightening discussions. A.D., H.K., G.A.D.-C. and L.S acknowledge the support of the Deutsche Forschungsgemeinschaft (DFG, German Research Foundation) -- Project-ID 274200144 -- SFB 1227 DQ-mat within the project A04, and under Germany's Excellence Strategy -- EXC-2123 Quantum-Frontiers -- 390837967. M.\L. acknowledges support from
the National Science Centre (Poland) via Opus Grant No. 2019/35/B/ST2/00838.
M.\L. and J.Z. gratefully acknowledge Polish high-performance computing infrastructure PLGrid (HPC Centers: ACK Cyfronet AGH) for providing computer facilities and support within computational grant no. PLG/2025/018457.
J.Z. acknowledges support by the National Science Centre (Poland)  under the OPUS call within the WEAVE program 2021/43/I/ST3/01142. 
The research has also been supported by the Priority Research Area (DigiWorld) under the Strategic Programme Excellence Initiative at Jagiellonian University (M.\L., J.Z.). 

\bibliography{bibliography_trlad}

\cleardoublepage

\setcounter{equation}{0}
\setcounter{figure}{0}
\renewcommand{\theequation}{S\arabic{equation}}
\renewcommand{\thefigure}{S\arabic{figure}}

\section{Supplemental Material}

\paragraph{Order parameters.--} For the case of itinerant dipoles~(EBHM), we distinguish between the different phases using the following order parameters. 
\begin{itemize}
\item Second derivative of the momentum distribution $\left.d^2 n(q)/dq^2\right|_{q=0}$
- negative in the SF phase, due to the maximum at $q=0$. 
\item Long-range chirality-chirality correlations $\kappa^2$ - finite in the CSF phase, vanishes in all other phases.
\end{itemize}    

\begin{figure}[b!]
\centering
\includegraphics[width=0.9\columnwidth]{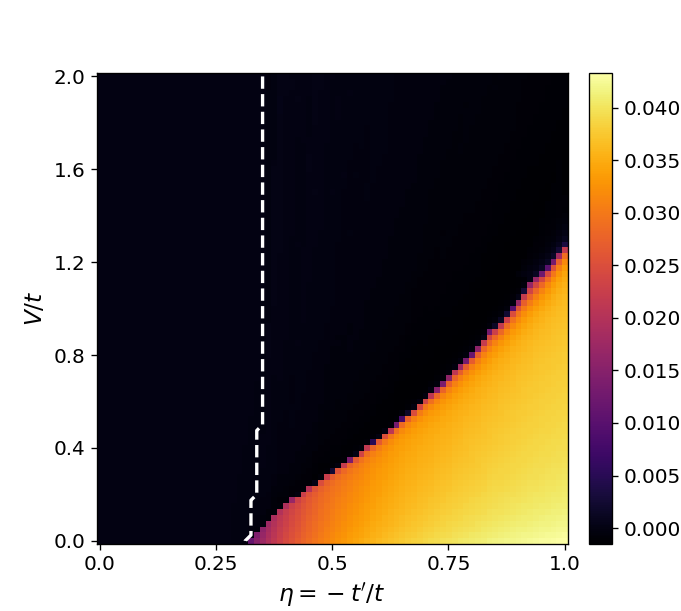}
\caption{Long-range chirality-chirality correlation $\kappa^2$ for the case of itinerant dipolar bosons. The white dashed line is the contour line $\left.d^2 n(q)/dq^2\right|_{q=0} = 0$, separating the SF phase from the other superfluid phases.}
\label{sup_1}
\end{figure}

Fig.~\ref{sup_1} shows our DMRG results for the two order parameters used to generate the phase diagram of Fig.~\ref{fig:2}(a). The colormap shows the long-range chirality-chirality correlations $\kappa^2$, while the white dashed line indicates the contour $\left.d^2n(q)/dq^2\right|_{q=0}=0$, which separates the 2SF and the SF phases. The calculations were carried out on a system of $L=120$ sites taking inter-site couplings up to two neighbours. 

For the case of pinned dipolar spins~(dipolar XXZ model) we characterize the phases using the following order parameters (in addition to the long-range chirality-chirality correlation $\kappa^2$):
\begin{itemize}
\item Nearest-neighbour pairing: $\Omega_{PSF}=\frac{1}{N}\sum \langle\hat S^z_i \hat S^z_{i+1}\rangle$  with $N=(1/2+\langle \hat S^z\rangle)L$ the total number of magnons in the system~($L$ denotes the number of sites). $\Omega_{PSF}$ presents a large value in the quadratic nematic phase~(i.e. for a pair-superfluid of magnons), indicating the proliferation of pairs on nearest-neighbour magnons. We have also checked that in the nematic phase single-magnon correlations decay exponentially, whereas 
single bi-magnon correlations decay polynomially. Further, we have confirmed that these bi-magnon nematic correlations are, for our range of parameters, always dominant compared to the longitudinal spin correlations $\langle \hat S_i^z \hat S_{i+r}^z \rangle - \langle \hat S_i^z \rangle\langle\hat S_{i+r}^z \rangle$, excluding a spin-density wave~(SDW$_2$) phase~\cite{Hikihara2008, Sudan2009, Heidrich-Meisner2010}

\item Inter-leg correlations: $\frac{1}{N}\sum_i \langle\hat S^+_i \hat S^-_{i+1}\rangle$. They are strongly suppressed in the 1DSF phase, as the two legs are effectively cut-off from each other. 
\item Standard deviation of the magnon momentum-distribution with respect to $q=\pi$:  $\Delta_q = \sqrt{\langle (q-\pi)^2\rangle}$. It is close to zero in the AFM-XY phase, nonzero in other phases.
\item Variance of the local occupation number of magnons: $\Delta_n =\frac{1}{L} \sum_i\langle\hat n_i^2\rangle-\langle\hat n_i\rangle^2$, with 
$\hat n_i = \frac{1}{2}+\hat S_i^z$. 
It is small in the SB-SDW phase as compared to the TLL phases.
\end{itemize}


\begin{figure*}[t!]
\centering
\includegraphics[width=\textwidth]{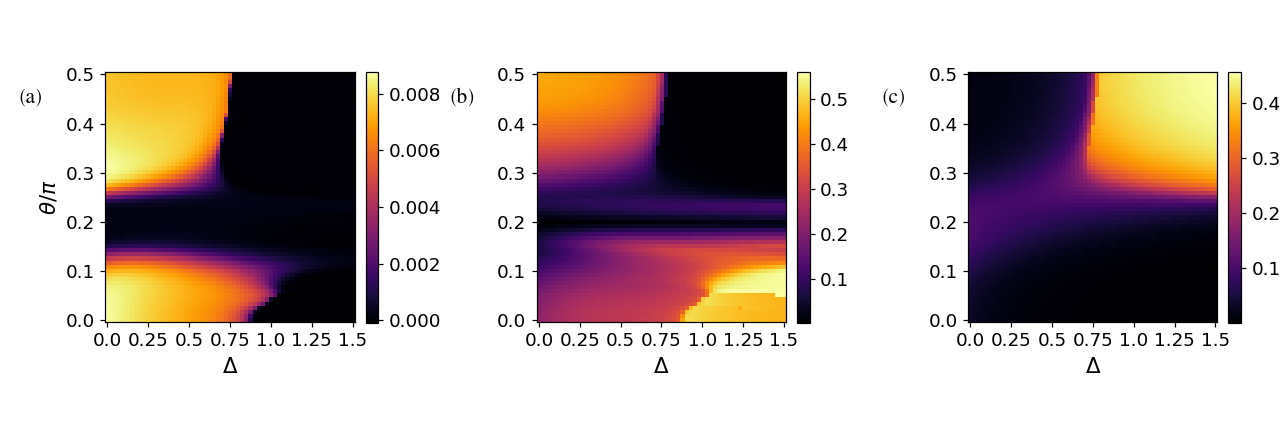}
\caption{Order parameters for the 
case of the dipolar XXZ model with the electric field on the $yz$ plane at magnetization $\langle \hat S^z\rangle = -0.4$: (a) long-range chirality-chirality correlation $\kappa^2$; (b) inter-leg correlations $\frac{1}{N} \sum_i \langle \hat S^+_i \hat S^-_{i+1}\rangle$; and (c) nearest-neighbor pairing, $\Omega_{PSF}$.}
\label{sup_2}
\end{figure*}



\begin{figure*}[t!]
\centering
\includegraphics[width=\textwidth]{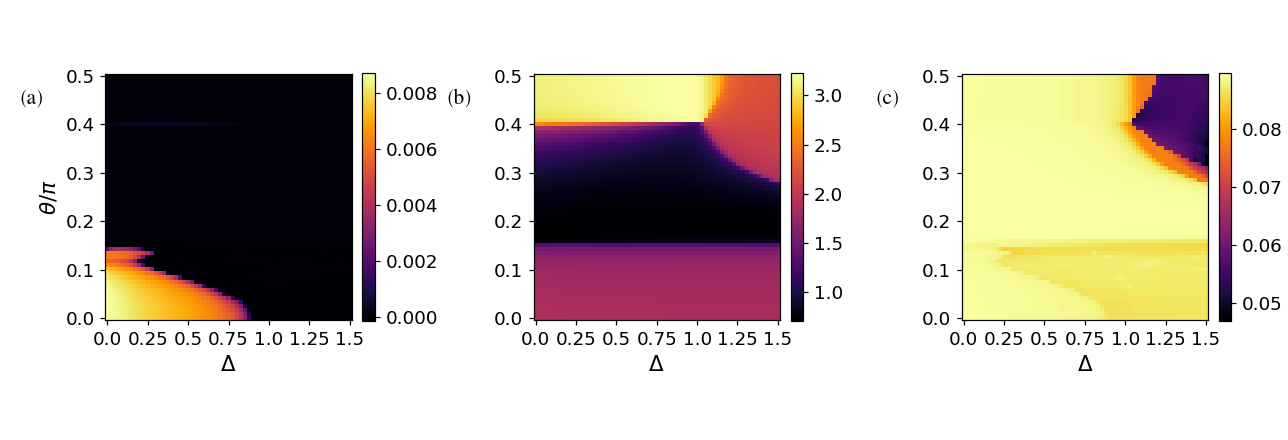}
\caption{Order parameters for the 
case of the dipolar XXZ model with the electric field on the $xz$ plane at magnetization $\langle \hat S^z\rangle = -0.4$: (a) long-range chirality-chirality correlation $\kappa^2$; (b) standard deviation in the magnon momentum with respect to $\pi$, $\Delta_q$; and (c) variance of the local occupation number of magnons, 
$\Delta_n$.}
\label{sup_3}
\end{figure*}



\begin{figure*}[t!]
\centering
\includegraphics[width=0.8\textwidth]{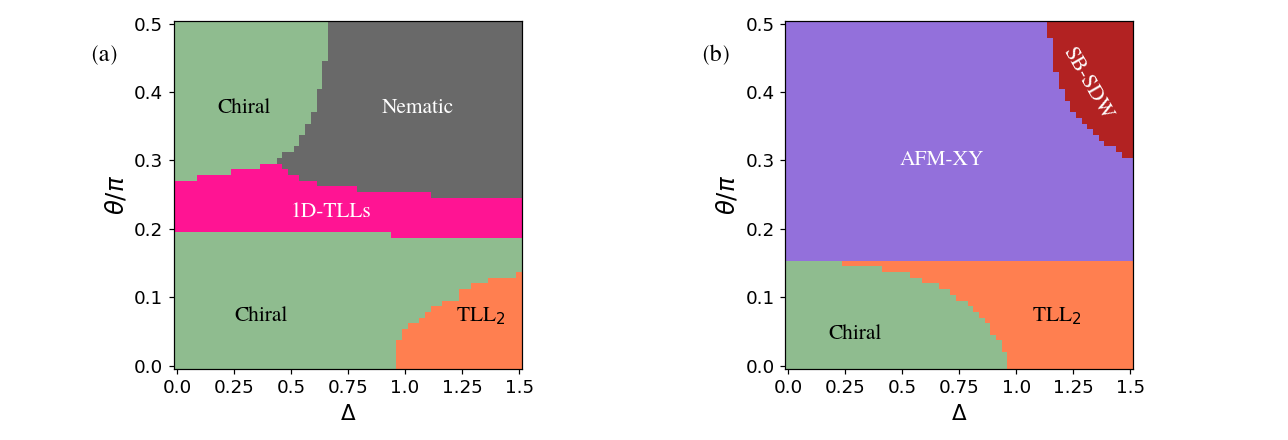}
\caption{Phase diagram for the pure 
$J_1$--$J_2$ model, for a  magnetization $\langle \hat S^z\rangle = -0.4$, for dipoles aligned at an angle $\theta$ with the $z$-axis, on the $yz$ plane~(a), and on the $xz$ plane~(b).}
\label{sup_4}
\end{figure*}

Figs.~\ref{sup_2} and~\ref{sup_3} show, respectively, the order parameters used to draw the phase 
diagrams of Figs.~\ref{fig:3}~(b) and (c). The calculations were performed using DMRG for the corresponding model of hardcore magnons with maximum bond-dimension of $200$ on a system of $120$ sites. In order to take the long-range hoppings into account, the calculations were carried out with inter-site couplings up to ten neighbours.

Our calculation for both the itinerant dipoles, and for the dipolar spins were performed using open boundary conditions. This was chosen in this way because on one hand it matches better with experimental conditions, and on the other hand working with periodic boundary conditions necessarily imposes a constraint in the possible rotation pitch of the phase in  the CSF, or spin orientation in the chiral spin phases.

\paragraph{Effect of the Dipolar Tail.--} Although the tail of the dipolar interaction $J_{r>2}$ does 
not alter the qualitative nature of most of the phases, it does result in significant shifts of the phase boundaries compared to the pure $J_1$--$J_2$ model in which $J_{r>2}$ are set to zero. In Figs.~\ref{sup_4}(a) and (b) we depict, respectively, the same phase diagrams as in Fig.~\ref{fig:3}~(b) and (c), but 
setting $J_{r>2}=0$. Note the shift in the boundaries. Note as well that, in the case of dipoles on the $xz$-plane, positive long-range hopping strengths from the dipolar tail lead to a dispersion with a global minimum at $q=0$ for large dipolar angles $\theta$, thus introducing a FM-XY phase which is absent in the pure $J_1$--$J_2$ model.
\end{document}